\documentclass[fleqn,12pt,twoside]{article}
\usepackage{graphicx}
\newcommand{\be}{\begin{eqnarray}}
\newcommand{\ee}{\end{eqnarray}}

\title{Locating strongly coupled color superconductivity
using universality and data for trapped ultracold atoms}

\author{  E.V. Shuryak \\
Department of Physics and Astronomy, \\ University at 
Stony Brook NY 11794-3800, USA
}
\begin{document}
\maketitle

\begin{abstract}
Cold fermionic atoms are known to enter the  universal
strongly coupled regime as their scattering length $a$ gets
 large compared to the inter-particle distances. Recent 
experimental data provide important critical
parameters of such system. We argue that quarks may enter
the same regime due to  marginal binding
of diquarks, and if so one can use its 
universality in order to deduce such properties as
the slope of the critical line of color superconductivity,
 $dT_c/d\mu$. We further
discuss limitations on the critical temperature itself
and conclude that it is limited by $T_c<70\, MeV$.
 \end{abstract}

1.{\em Color superconductivity} (CS) in dense quark matter
in general appears due to attractive interaction in certain scalar
diquark channels. More specifically, three mechanisms
of such attraction were discussed in literature
are: (i) the electric Coulomb interaction 
(see e.g.\cite{Bailin:1984bm}) 
(ii) instanton-induced 't Hooft interaction 
\cite{RSSV,ARW} and the 
magnetic interaction \cite{Son:1999uk}.

Unlike electrons in ordinary superconductors
(for which the driving attraction is a complicated exchange of
collective excitations like
phonons), quarks naturally may have  color charges leading to
 their mutual electric  attraction.
However this straightforward mechanism of CS
 is significantly weakened by the
electric screening, at the ``Debye mass'' scale
$M_D\sim g\mu$. It thus leads to
 rather weak pairing,  with the gaps 
 $\Delta\sim 1\, MeV$ \cite{Bailin:1984bm}.

The second mechanism of pairing is due to  the {\em instanton-induced}
forces between light quarks, also known as the 't Hooft interaction.
It is related to fermionic zero mode of instantons and thus it 
is very flavor-dependent. In particularly, it only exists
for two quark  of different flavors (such as $ud,us$ and $ds$).
 Therefore
the simplest setting in which it was considered is
the  2-flavor (no $s$-quark)
problem  \cite{RSSV,ARW}, for 3-flavor case see \cite{RSSV2}. The
instanton mechanism have re-activated the field, because it
 have demonstrated for the first time
 that large (and thermodynamically significant)
pairing may actually appear in cold quark matter. Since
forces induced by small-size instantons are 
not affected by screening that much, they can be much stronger
than (i), leading to
 gaps  about 2 orders of magnitude larger
 $\Delta\sim 100\, MeV$ \cite{RSSV,ARW}. 

  However such gaps are rather uncertain because
 they are
so large\footnote{ 
Additional argument \cite{RSSV} why one should believe such large gaps:
 the $same$ interaction but in $\bar
q q$ channel is responsible for chiral symmetry breaking, producing the gap
(the constituent quark mass) as large as 350-400 MeV.
Furthermore, in the {\it two-color} QCD, the so called Pauli-Gursey
symmetry  relates these two condensates directly.} that they
may fall outside of applicability range
of  the usual BCS-like mean field theory. Unfortunately,
an analytic theory of  strong pairing is still
in its infancy.
 The main idea of this letter
is  to get around this 
``calculational'' problem by using the universality arguments
and  an appropriate atomic data.

Magnetic  mechanism of pairing (iii) has been
 pointed out by Son \cite{Son:1999uk}: it dominates at very high
densities ($\mu> 10 GeV$) when two others  get strongly screened.
It
leads to  gaps 
$ \Delta \sim \mu exp( -{3\pi^2 \over \sqrt{2}g(\mu)})$, large
 in absolute
 magnitude but small relatively to $\mu$.

  Different colors and flavors of quarks lead to
  to multiple possible condensates, and indeed
color superconductivity may exist in many different phases
depending on $T,\mu$ and quark masses.
In particular, 
they may have chiral and translational symmetries being either
preserved or broken. 
We will not go into this vast subject: the reader
for definiteness may think about a single $ud$ scalar
condensate, or the so called 2SC phase.
Its symmetries  are similar to
the 
electroweak part of the Standard Model, with the
fundamental color representation of the $ud$ condensate,
 breaking the color group and
making 5  gluons massive and leaving 3 (of the unbroken SU(2))
massless.

 The estimates for the {\em maximal pairing} possible
are obviously very interesting for applications.
One of them is a possibility that a superconducting quark matter is
present at the central region of compact ``neutron''
stars.
Another is a possibility that either a region
of the CS phase on the phase diagram,
or at least a region where diquarks are bound
can be reached via heavy ion collisions. In both cases
the issue of the gap magnitude (or critical temperature) is crucial
for making those possibilities real.

2. Let us now introduce
the main idea of this letter. The interaction (scattering) 
of two quarks
is maximally  enhanced (to its unitarity limit) if there is a marginal 
state in the diquark channel,   a bound state
with near-zero binding or  a virtual state at small positive
energy. In atomic systems such situations are generically called
a ``Feshbach resonance'', tuned by external magnetic field.
 Transition between these two possibilities is 
reflected in specific behavior of the condesation, known as
BEC-to-BCS transition. In the middle, with resonance at
exactly zero binding, the interaction is
at its maximum, limited by
unitarity for the relevant partial wave. (It is s-wave with l=0
in all cases considered.)

In quark-gluon plasma  the existence of marginally bound states
of quarks and gluons 
and their possible role in liquid-like behavior at RHIC has been
pointed out in \cite{SZ_1}. 
For any hadronic states 
one can argue that they dissolves at large $T$ or $\mu$,
and thus existence of some lines of zero binding
on the phase diagram are unavoidable.
The issue was so far studied only for small
$\mu$ and high $T$ relevant for RHIC heavy ion program:
and indeed there are theoretical, lattice and experimental
evidences that e.g. charmonium ground state does not melt till
about $T\sim 400\, MeV$ \cite{SZ_2,charmonium}. One can also follow  discussion of
such lines for glueball and light quark mesons in 
\cite{SZ_2}, charmed mesons in \cite{Rapp:2005at},  baryons and multi-gluon chains
in \cite{Liao}. 

 In this letter we explore consequences of the idea that
there is a $diquark$ marginal binding line, more or less
trailing the phase boundary line on the phase diagram.
If this is the case (which cannot be proven at this time),
it must cross the CS critical line, separating the CS region
into a BCS-like and BEC-like.  
 (For general information
on BEC-BCS transition and its possible presence in quark matter
see e.g. a lecture \cite{BEC-BCS_q}.)

\begin{figure}[h]
\includegraphics[width=5.5cm]{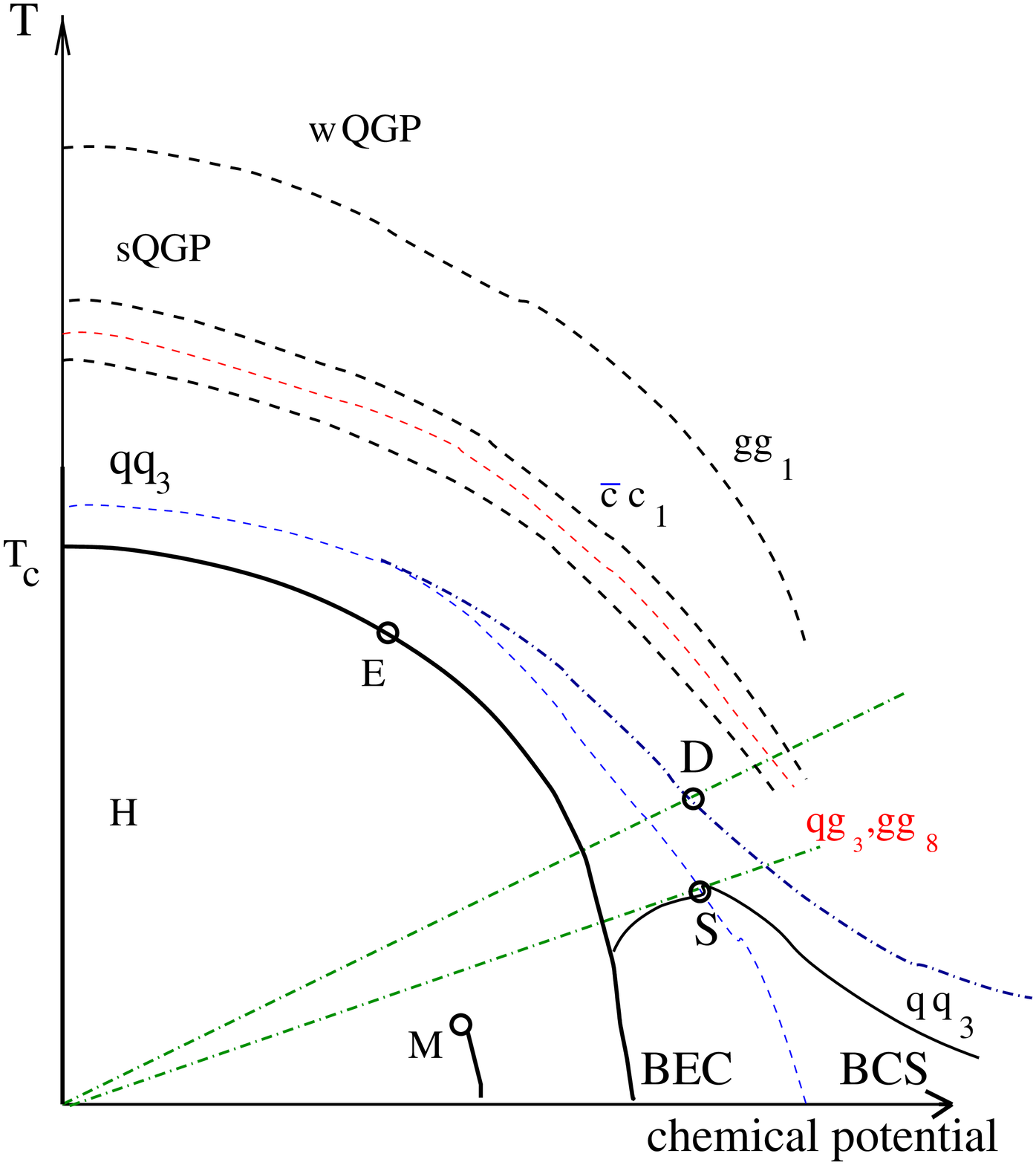}
\hfill
\includegraphics[width=7cm]{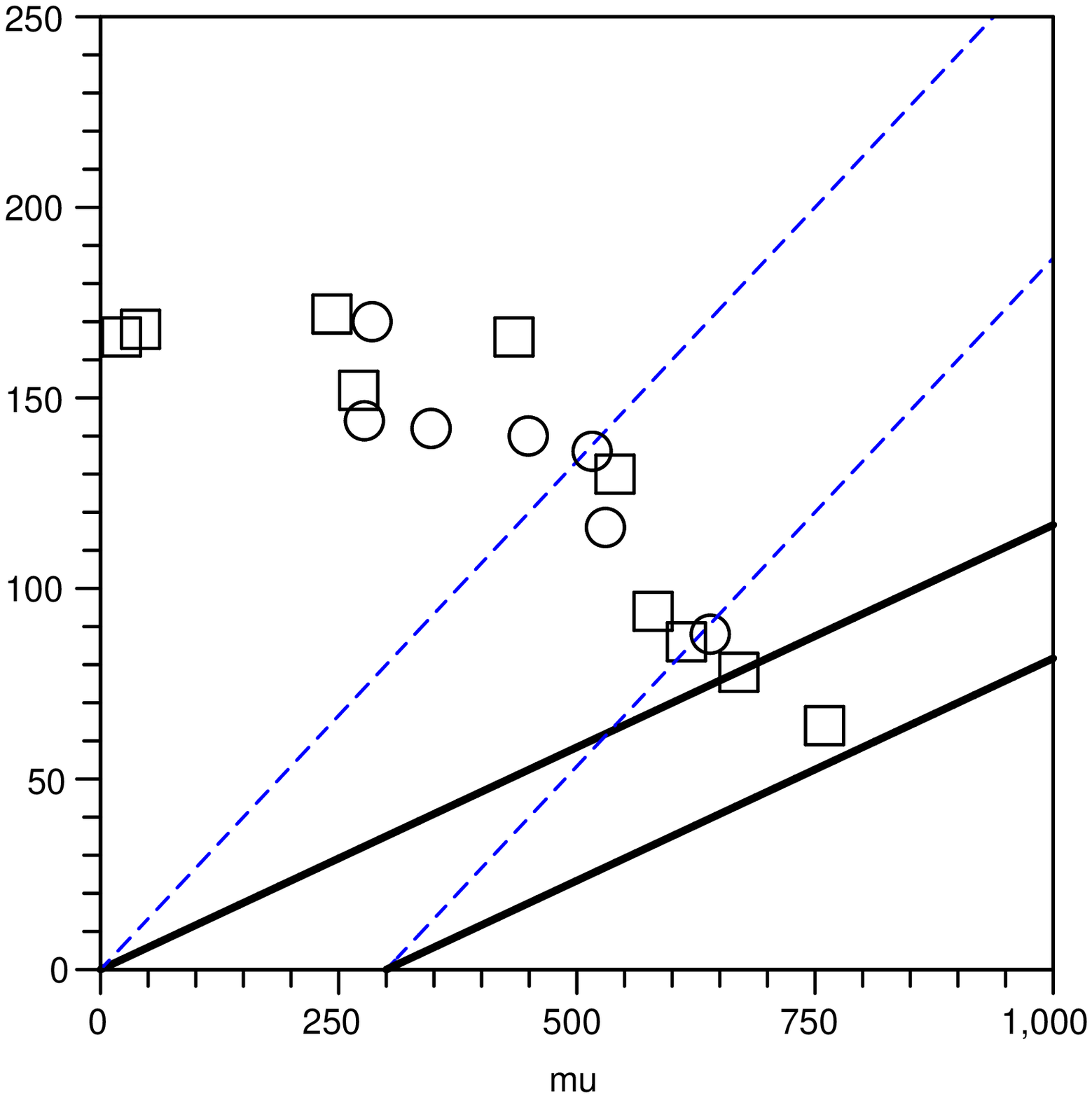}
\caption{(a) Schematic phase diagram for QCD, in the plane baryon
  chemical potential - temperature.
M (multifragmentation) point is the endpoint of nuclear gas-liquid transition.
E is a similar endpoint separating the first order transition
to the right from a crossover to the left of it.
(Black) solid lines show
phase boundaries, dashed  lines are curves
of marginal stability of indicated states. 
Two dash-dotted straight lines are related with bounds from atomic
experiments we discuss in the text, they intersect with
unbinding of diquark Cooper pairs (D) and most strongly coupled point (S),
which is at the maxumum of the transition line ans is
also a divider between BCS-like and BEC-like color superconductor.
(b) Compilation of experimental data on chemical freezeout parameters
from different experiments according to
\protect\cite{chemical}.
The squares and circles are for fits at mid-rapidity and all
particles, respectively. Two solid lines are the phase transition lines
with the quark effective mass $M_1=0$ and $100\, MeV$, two dashed
  lines show pair unbinding lines for
 the same masses.
}
\label{fig_phase_diag}
\end{figure}

In Fig.\ref{fig_phase_diag}(a) we show schematic phase diagram
of QCD, in coordinates baryonic chemical potential $\mu$ (per quark)-
temperature $T$. Lines of zero binding\footnote{Similar
 ``curves of marginal  
stability''  (CMS) of certain states
 are known in various settings: e.g. they have an important role
in confinement of supersymmetric gauge theories \cite{RV}.} of a $qq$ pair \cite{SZ_1} 
 starts at $\mu=0$ at the temperature $T_{qq}$ very close to
the critical line $|T_{qq}-T_c|\ll T_c$: the reason for that
is that effective color attraction in $qq$ state is only $1/2$
of that in mesons such as charmonium. 
 The lines associated with (color triplet) diquark
$qq_3$ are
the
lower (blue) dashed and dash-dotted lines: below the former  
diquark binding is below zero and above the later it does not exist at
all,
even as a Cooper pair.

3.
How can one relate  atomic and  quark systems, if at all?
The way it can be done is due to the so called ``universality''
of the system. As the scattering length gets large $a\rightarrow
\infty$,
it cannot enter the answers any more. As a result, 
there remain so few parameters on which the answer may
depend\footnote{
The next parameter of scattering amplitude, the effective range,
is about 3 orders of magnitude smaller than interparticle distance
for trapped atoms, and thus completely irrelevant.
Although similar parameter for
quarks are not that small, we will assume it does not
affect the universal results too much.
}
on that those  can be absorbed by selection of the appropriate units.

Atomic 50-50 mixture of two ``spin'' states is characterized
by the density $n$ and the atomic mass $m$. 
Quantum mechanics adds $\hbar$ to the list of possible quantities,
and so one has only 
3 input parameters, which can be readily absorbed by
selecting proper units of length, time and mass. Thus the pressure
(or mean energy) at
infinite and zero $a$ can
only
be  related by some universal
 numerical  constant 
 $p_\infty / p_0= (1+\beta)$. We do not know how to get 
its value from any theory
(other than quantum Monte-Carlo simulations or other brute force
methods), 
 but it has been  measured 
experimentally (e.g. by the very size of the trapped system). 
The same should hold for transport properties: e.g.
viscosity of such universal gas
 can only be  $\eta=\hbar n \alpha_\eta$ where $\alpha_\eta$
is some universal coefficient \cite{GSZ}. 
Similarly, the critical temperature 
must be simply proportional to the Fermi energy 
\be \label{eqn_alpha_def}
T_c=\alpha_{T_c} E_F\ee
with the universal constant $\alpha_{T_c}$.

Experimental progress in the field of strongly coupled
 trapped fermionic atoms is quite spectacular; unfortunately
 this author is certainly unqualified to  go into its discussion.
 Let me just mention one paper, which killed 
remaining doubts about superfluidity: 
 a discovery by the MIT group of
the quantized vortices, neatly
organized
into the usual lattice
\cite{MIT_vort}.
 We will however need only  the
  information about the $value$ of the transition temperature. 
Duke group lead by J.E.Thomas
(Kinast et al) have for some time studied collective vibrational
modes of the trapped system. Their frequencies in strong coupling
regime are well predicted by hydrodynamics and universal
equation of state, with little variation with temperature.
Their dampings however show
significant $T$-dependence: in fact Kinast et al
\cite{kinast_damping} have found two distinct transitions
in its behavior. 
The lowest break in damping
is interpreted as the phase transition to superfluidity, it
corresponds to \be \alpha_{T_c}={T_c\over E_F}=.35\ee
 where $E_F$ stands for
the Fermi energy of the ideal Fermi gas at the center of the
trap.
 In another (earlier) set of experiments
 \cite{kinast_cv} there has also been found a change in the specific
heat,   at   $\alpha_{T_c}=.27$. In spite of some
numerical difference between these two values,
 the Duke group indicates
that both are related to the same phenomenon\footnote{ They
  do not provide any
error bars on the value of $T$, as the absolute value of the temperature
is obtained in rather indirect calibration procedure. The reader
may take the spread of about 20\% as an estimate of uncertainties
involved.}. 
At another  temperature  
\be \alpha_2={T_2\over T_F} \approx 0.7-0.8 \ee
 the behavior of the damping
visibly changes again. Kinast et al
 interpret it as a transition to a regime where
not only there is no $condensate$ of atomic pairs, but
even the  pairs themselves are melted out.

4. 
There are of course important differences between quarks and atoms.
First, quarks have not only spin but also flavor and color, so
there are $3*N_f$ more Fermi surfaces. However, in the first approximation one
may focus only at one pair of them (say u-d quarks with red-blue
colors)
which are actually paired.
Second, atoms are non-relativistic while quarks are 
in general relativistic 
and in matter may have some complicated dispersion laws.
Since the gaps are large, one cannot use a standard argument
that close to Fermi surface only Fermi velocity is important.
Nevertheless, we will
 assume that quark quasiparticles
have dispersion laws which can be approximated by a simple
quadratic form
\be E(p)=M_1+{p^2\over 2 M_2}+... \ee 
where dots are for $O(p^4)$ terms we ignore.  In matter
$M_1$ and $M_2$ need not be the same. 
 It then implies that e.g. the relation between the critical $T$ and
chemical potential
should read
\be  T_c= \alpha_{T_c}(\mu/3-M_1)\ee
and similarly for the second point.
(The factor 3 appears because baryon number is counted per baryon, not quark.)

The $M_2$ can be used
to set the units as explained above, 
and we will not actually need its value.
 The $M_1$ is  needed, but since it makes   a simple shift of
the chemical potential, it can be eliminated by differenciation.
Thus we get a prediction of the slope of the critical
line 
\be dT_c/d\mu= \alpha_{T_c}/3\approx 0.1\ee  
The intersection of CMS for diquarks
with the SC critical line (the point S (strong) in our phase diagram   
Fig.\ref{fig_phase_diag}(a)) should
thus be at this line, at which the boundary of superconducting
phase
crosses with the zero energy of the bound state.
The second critical point associated with disappearance of pairs
(identified with the (blue) dash-dotted line and point D
in Fig.\ref{fig_phase_diag}(a)) should thus be at the line with the
slope
$\alpha_2$. 

In order to plot the line on the phase diagram one needs the value
of the $M_1$, which unfortunately is not known. To set the
 {\em upper bound}
one may simply take $M_1=0$ and draw
two straight lines pointing to the origin,
see Fig.1(b). 
As $M_1$ grows, the lines slide to the right,
as is shown by another line with a
 (randomly chosen) value $M_1= 100 \, MeV$.

Finally, we turn to ``realistic'' phase diagram, with
numerical values extracted from experiment.
 We know for sure  that matter is released at
the so called chemical freezeout lines indicated by points in
Fig.\ref{fig_phase_diag}(b). These points are
the ends of adiabatic cooling paths.  Unfortunately we do not
know how high above them those curves start at a given collision
energies. However it is general expected that this line more or less
traces the critical line, being few MeV below it, into
 the hadronic phase. One may then conclude that the upper limit
on $T_c$ of CS is about 70 MeV
(intersection of the upper solid lines with the freezeout line). The
disappearance of pairing (dashed lines) is thus expected below
$T_2=150\, MeV$. 

Finally, comparing these results with theoretical expectations
and experimental capabilities, we conclude that  
(i) if there is a strongly coupled CS,
its critical temperature should definitely be below $T_c<70\, MeV$.
This maximal value is amusingly close  to
what was obtained from 
the instanton-based calculations \cite{RSSV,ARW};
(ii) it is unlikely that any heavy ion collisions
can reach the CS domain, 
even at the maximal coupling. A penetration into the region
$T= 100-150\, MeV, \mu=500-600\, MeV$ in which
non-condensed bound diquarks may exist, is however quite likely,
both in the low-energy RHIC runs and in future GSI facility FAIR;
(iii) if that happens, one may think of some further
 uses of universality,
e.g. about relating the transport properties in both systems.
One may in particularly ask whether the universal
 viscosity extracted from vibrations
of trapped atoms (like that
in \cite{GSZ}, but at appropriate $T$)
 can or cannot describe the hydrodynamics
of the corresponding heavy ion collisions.

{\bf Acknowledgments.}
This work was initiated at the Trento meeting related to GSI
future facility FAIR in June 06: I thank the organizers for
forcing me to thing about the issues discussed in this letter.
It is  also partially supported by the US-DOE grants DE-FG02-88ER40388
and DE-FG03-97ER4014.


\begin{thebibliography}{99}
\bibitem{Bailin:1984bm}
D.~Bailin and A.~Love,
Phys.\ Rept.\  {\bf 107}, 325 (1984).
\bibitem{RSSV}
R.~Rapp, T.~Sch{\"a}fer, E.~V.~Shuryak and M.~Velkovsky,
Phys.\ Rev.\ Lett.\  {\bf 81}, 53 (1998)
[hep-ph/9711396].

\bibitem{ARW}
M.~Alford, K.~Rajagopal and F.~Wilczek,
Phys.\ Lett.\  {\bf B422}, 247 (1998),
[hep-ph/9711395].

\bibitem{Son:1999uk}
D.~T.~Son,
Phys.\ Rev.\  {\bf D59}, 094019 (1999)
[hep-ph/9812287].
\bibitem{RSSV2}
R.~Rapp, T.~Sch\"afer, E.~V.~Shuryak and M.~Velkovsky,
{\it Ann. Phys.\ }  {\bf 280}, 35 (2000), hep-ph/9904353.

\bibitem{SZ_1}E.~V.~Shuryak and I.~Zahed,
  Phys.\ Rev.\ C {\bf 70}, 021901 (2004)
  [arXiv:hep-ph/0307267].
\bibitem{SZ_2}E.~V.~Shuryak and I.~Zahed,
Phys.\ Rev.\ D {\bf 70}, 054507 (2004)
  [arXiv:hep-ph/0403127].
\bibitem{charmonium}
  M.~Asakawa and T.~Hatsuda,
  Phys.\ Rev.\ Lett.\  {\bf 92}, 012001 (2004)
  [arXiv:hep-lat/0308034].
S.~Datta, F.~Karsch, P.~Petreczky and I.~Wetzorke,
  Phys.\ Rev.\ D {\bf 69}, 094507 (2004)
  [arXiv:hep-lat/0312037].
\bibitem{Rapp:2005at}
  R.~Rapp, V.~Greco and H.~van Hees,
  arXiv:hep-ph/0510050.
\bibitem{Liao}
  J.~Liao and E.~V.~Shuryak, Nucl.Phys.A, in press,
  arXiv:hep-ph/0508035.
Phys.\ Rev.\ D {\bf 73}, 014509 (2006)
  [arXiv:hep-ph/0510110].
\bibitem{BEC-BCS_q}B.Kerbikov, hep-ph/0510302
\bibitem{RV} 
A.Ritz, M.~A.~Shifman, A.~I.~Vainshtein and M.~B.~Voloshin,
  Phys.\ Rev.\ D {\bf 63}, 065018 (2001)
  [arXiv:hep-th/0006028].

\bibitem{MIT_vort}M. W. Zwierlein, J. R. Abo-Shaeer, A. Schirotzek, C. H. Schunck and W. Ketterle,Nature 435, 1047-1051 (23 June 2005) 


\bibitem{GSZ}  B.~A.~Gelman, E.~V.~Shuryak and I.~Zahed,
Phys. Rev. A 72, 043601 (2005), nucl-th/0410067.



\bibitem{kinast_cv} J. Kinast et al, Science 25, jan.2005
\bibitem{kinast_damping}J. Kinast et al, cond-mat/0502507)
\bibitem{chemical}
A. Andronic, P. Braun-Munzinger, J. Stachel, 
Nucl.Phys.A772:167-199,2006,
 nucl-th/0511071
\end{thebibliography}
\end{document}